\documentclass[sigconf]{acmart}
\AtBeginDocument{%
  \providecommand\BibTeX{{%
    \normalfont B\kern-0.5em{\scshape i\kern-0.25em b}\kern-0.8em\TeX}}}

\setcopyright{ccbysa} 
\copyrightyear{2022}
\acmYear{2022}

\acmConference[CHI '22]{CHI Conference on Human
Factors in Computing Systems}{April 29--May 5, 2022}{New Orleans, LA, USA}
\acmBooktitle{CHI Conference on Human Factors in Computing Systems (CHI '22), April 29--May 5, 2022,
New Orleans, LA, USA}

\usepackage{color}
\usepackage{hyperref}
\usepackage{caption}
\usepackage{subcaption}

\begin{document}

\title{Data-Driven Visual Reflection on Music Instrument Practice}


\settopmatter{authorsperrow=1}


\author{Frank Heyen}
\email{frank.heyen@visus.uni-stuttgart.de}
\orcid{0000-0002-5090-0133}
\author{Quynh Quang Ngo}
\email{quynh.ngo@visus.uni-stuttgart.de}
\orcid{0000-0001-5254-1480}
\author{Kuno Kurzhals}
\email{kuno.kurzhals@visus.uni-stuttgart.de}
\orcid{0000-0003-4919-4582}
\author{Michael Sedlmair}
\email{michael.sedlmair@visus.uni-stuttgart.de}
\orcid{0000-0001-7048-9292}
\affiliation{%
  \institution{VISUS, University of Stuttgart}
  \city{Stuttgart}
  \country{Germany}}

\renewcommand{\shortauthors}{Heyen, et al.}

\begin{abstract}
We propose a data-driven approach to music instrument practice that allows studying patterns and long-term trends through visualization.
Inspired by life logging and fitness tracking, we imagine musicians to record their practice sessions over the span of months or years.
The resulting data in the form of MIDI or audio recordings can then be analyzed sporadically to track progress and guide decisions.
Toward this vision, we started exploring various visualization designs together with a group of nine guitarists, who provided us with data and feedback over the course of three months.
\end{abstract}



\begin{teaserfigure}
     \centering
     \begin{subfigure}[b]{0.24\linewidth}
         \centering
         \includegraphics[width=\linewidth]{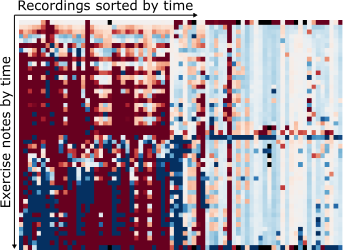}
         \caption{
         Progress over multiple recordings.
         While improving overall, issues with some notes remain.
         }
         \label{fig:teaser_progress}
         \vfill
     \end{subfigure}
     \hfill
     \begin{subfigure}[b]{0.24\linewidth}
         \centering
         \includegraphics[width=\linewidth]{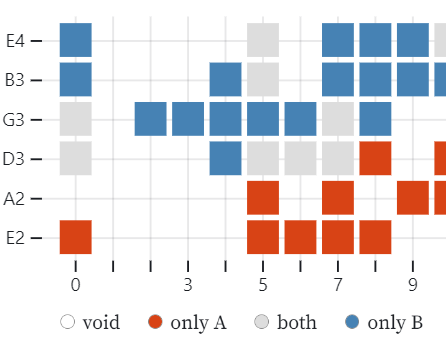}
         \caption{
         A fretboard heatmap shows which notes were played by two guitarists.
         }
         \label{fig:teaser_heatmap}
     \end{subfigure}
     \hfill
     \begin{subfigure}[b]{0.24\linewidth}
         \centering
         \includegraphics[width=\linewidth]{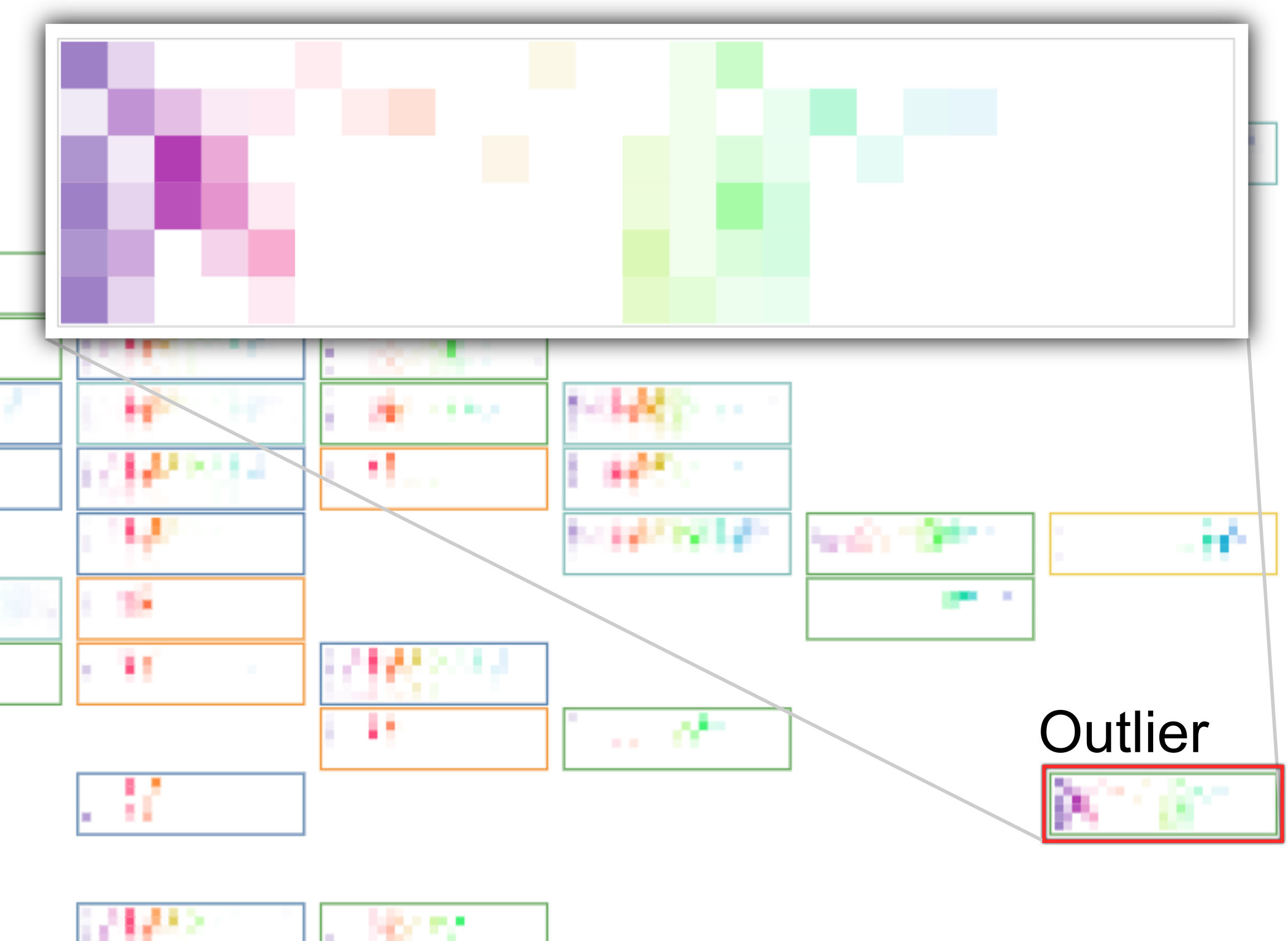}
         \caption{
         Multiple heatmaps positioned by similarity. 
         Colors allow quickly detecting differences.
         }
         \label{fig:teaser_heatmap_dr}
     \end{subfigure}
     \hfill
     \begin{subfigure}[b]{0.24\linewidth}
         \centering
         \includegraphics[width=\linewidth]{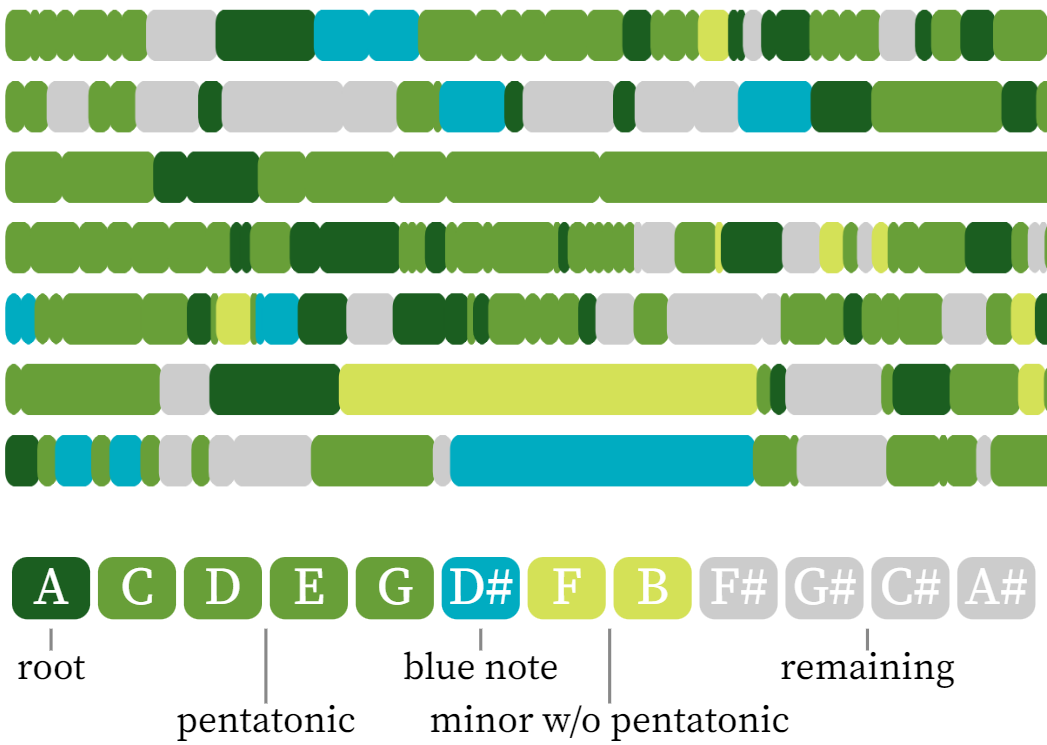}
         \caption{
         Recorded notes colored by their role for improvisation.
         Each row shows one recording.
         }
         \label{fig:teaser_scales}
     \end{subfigure}
  \caption{
   Four examples of our visualization designs.
    }
  \Description{
  Four example images of our visualization designs. 
  From left to right:
  a)
    A heatmap with x axis for recordings and y axis for the notes of the exercise.
    Each cell stands for the error of a note in the corresponding exercise.
    Colors get lighter from left to right, indicating improvement over time.
    Some darker horizontal lines show notes that are still not played correctly even in later recordings.
  b)
    A fretboard heatmap has an x axis for the guitar's frets and a y axis for its strings.
    Each cell stands for a note on the fretboard and cells are colored, in this case, either red, when only player A played it, blue, when only player B played it, gray, if both played it, or transparent, if none played it.
  c)
     Multiple fretboard heatmaps are placed in a grid, more similar looking ones are closer to each other.
     In the bottom-right, there is a single heatmap marked as an outlier, a zoomed in version of it is shown at the top.
     Each heatmap is colored in a rainbow scheme to give each fret (column) a different color so similarities and differences between heatmaps become more apparent.
  d)
     Multiple rows of colored rectangles.
     The bottom most row has a large blue rectangle.
     Below, there is a legend that shows what colors are mapped to what note, for example the A minor pentatonic scale is green and the blue note blue.
  }
  \label{fig:teaser}
\end{teaserfigure}

\maketitle

\section{Introduction}

Music has long been part of culture~\cite{morley2013prehistory, mithen2006singing} and positively influences well being~\cite{fink2021viraltunes}, which leads many to not only enjoy listening to music but also play instruments themselves. 
However, acquiring the necessary skills and knowledge demands spending time and effort on repetitive practice.
While teachers can assess skills and provide guidance, such immediate feedback is not accessible during the exercise between lessons.

Based on a similar motivation, work in other education domains, such as sports training~\cite{perin2018sports:data:vis, lin:chi2021ar:vis:basketball, shum2016skillvis}, has started to increasingly leverage data-driven approaches to expand this feedback loop with an additional data angle. 
Visual analysis can then be used to guide decisions and increase motivation~\cite{huang2015personal:vis:and:va}.
However, to the best of our knowledge, approaches that systematically record and visually analyze data for music education are still largely missing. 

We envision a similar data-driven learning process for music education that incorporates recorded play data from students and teachers. 
We believe that such approaches can further enrich --- but not fully replace --- traditional instrument teaching and learning processes.
Since music is diverse and often subject to taste and culture, automatic assessment cannot fully replace human judgement~\cite{eremenko2020performance}.
This calls for a human-in-the-loop solution as commonly provided by visual analysis interfaces~\cite{sacha2017you,sedlmair2012design}. 
Using data visualization, teachers and students can then analyze performance, skills, and progress while bringing in their experience and subjective judgement.
Let us take improvisation as an example: 
In this creative task, there is no clear and universal judgment for good or bad and automatic analysis might thus not be sufficient.
Instead, a heatmap visualization that shows how often notes on a guitar fretboard were played by each student would allow teachers to compare different improvisation styles and evaluate them based on their own expertise --- just as one example.

To start exploring this new design space, we conducted a design exploration study on data- and visualization-driven instrument learning by recording and analyzing data from nine guitar players over a span of three months.
This process resulted in different visualization designs, of which we present and discuss four in this paper (\autoref{fig:teaser}).

\section{Related Work}

Visualization with focus on music-related data covers a wide range of applications~\cite{khulusi2020survey}, 
including the analysis of raw audio~\cite{foote1999visualizing}, abstract meta data~\cite{baur2010streams}, repeating patterns~\cite{wattenberg2002arcdiagrams}, personal listening histories~\cite{baur2010streams}, and music theoretical aspects of pieces~\cite{miller2019framing, miller2019augmenting, fuerst2020augmenting}. 
As visual analysis of instrument-generated data for education purposes has not yet been addressed by research, we especially focus on amateur exercise recordings and visual comparison~\cite{gleicher2011visual} between different people's personal~\cite{huang2015personal:vis:and:va} recordings. 

Instrument education games such as
\emph{Rocksmith}\footnote{\url{https://rocksmith.ubisoft.com/}}, 
\emph{Yousician}\footnote{\url{https://yousician.com/}}, 
and \emph{Synthesia}\footnote{\url{https://synthesiagame.com/}} 
only provide immediate or coarsely aggregated feedback, mostly limited to simple scores or bar and line charts.
Moreover, automatic assessment does not yet meet the requirements of professional music education~\cite{eremenko2020performance}.
We therefore propose a user-centered approach where a teacher or student makes the assessment, supported by visualization on multiple levels of detail and abstraction.

Other work investigated human-computer interaction for musical instruments:
\emph{Strummer}~\cite{ariga2017strummer}, for instance, teaches playing guitar chords using audio input for assessment.
Colored sheet music~\cite{asahi2018toward:piano:support, hori2019piano:hmm, smith2008interactive} can visualize assessments by highlighting mistakes and comparing student's and teacher's recordings. 
Feedback can be provided on screens, but also through augmented reality displays for drum kits~\cite{yamabe2011feedback} and guitars~\cite{loechtefeld2011guitar}.
Instruments themselves can be fitted with touch sensors and LEDs to track exercises and show feedback~\cite{marky2021letsfrets}.
Sensor measurements were also used by some approaches to adapt the exercises' tempo or difficulty~\cite{karolus2018emguitar,yuksel2016bach} or let users control effects~\cite{karolus2020thumb}.
\emph{Soloist}~\cite{wang2021soloist} makes use of existing material by creating overviews of instructional videos for guitar and extracting the played notes from audio through CREPE~\cite{kim2018crepe}, so users can visually compare their playing to the instructor's in a simple chart.
The above approaches focus on live control or feedback for single recordings.
Instead, we want to show much larger data collections at once.

\section{Process \& Data}

The primary goal of our work is to start exploring the design space of data-driven music training. 
To that end, we opted for a data-driven, participatory design process~\cite{sedlmair2012design, hayes2011relationship, hutchinson2003technologyprobes}, in which we created and explored different visualization designs together with nine guitar players over a period of three months. 
As participants were located in different cities, we arranged weekly online meetings that we loosely structured as a guitar course with one teacher and eight students with different skill levels. 
This teacher previously taught guitar at a music school for seven years. 

The exercises given to the students consisted of (1) scale patterns, (2) riffs from actual songs, and (3) improvisations playing along an A-minor blues backing track.
Throughout the course, we analyzed recorded data together through descriptive and aggregated visualizations (see below). 
We used formative feedback from participants for iterative design adjustments.

In terms of data, we opted for MIDI~\cite{moog1986midi} over direct audio, as discrete note events are easier to interpret both by computers and humans. 
We used guitars fitted with Fishman MIDI pickups\footnote{\url{https://www.fishman.com/tripleplay/}} and a Jamstick MIDI guitar\footnote{\url{https://www.zivix.co/jamstik}}.
All participants were equipped with this hardware and asked to record their practice data whenever possible.

\section{Visualization Designs}

Since our design is not targeted at visualization experts, we designed rather simple visualizations and focused on aggregations and domain-specific representations, of which we describe four examples in the following subsections. 
\autoref{fig:teaser} shows screenshots of these examples.

\subsection{Comparison to Sheet Music \& Progress Tracking}

For most music pieces and instrument exercises, there is sheet music available that a student tries to follow closely while learning.
A comparison to this ``ground truth'' data therefore allows to estimate skills and progress.
As explained above, automatic approaches are not reliable in this task.
We therefore use visualization and aggregation to let a student or teacher \emph{take a look} and judge based on context and experience.

Our visualizations show different levels of detail, starting with an overview~\cite{shneiderman2003theeyeshaveitmantra} that shows the students' errors over all notes (top to bottom)  and all repetitions (left to right) of an exercise (\autoref{fig:teaser_progress}).
Each cell of the heatmap represents a note and is colored by the difference in timing between recording and sheet music: from dark blue for too early, over white, to dark red for too late.
In the shown example, colors get lighter with more repetitions  as the student improves, while there remain issues with some notes, indicated by horizontal stripes.
Alternative representations show these colors inside sheet music, either for a single recording or as a mean of multiple. 

\subsection{Instrument Heatmaps}

The following three examples use recordings of improvisation over a common backing track.
Here, players are not interested in following a music piece, but instead want to express themselves and find personal styles.
Comparison to other players, for example a teacher or a famous artist, can help learn new playstyles.

A simple way to summarize an improvisation is to use instrument heatmaps.
They show a simplified image of the physical instrument and encode information therein, for example by coloring instrument parts.
As we are using guitars, we designed different heatmaps showing the guitar's fretboard with strings as rows and frets as columns (\autoref{fig:teaser_heatmap}).
Each cell represents a note, positioned on a given string and fret, which can be colored by how often it was played in one or multiple recordings.
Looking at such heatmaps of different players allows comparing their general style.
In \autoref{fig:teaser_heatmap}, for example, we can see that player \emph{A} (red) uses the lower strings, but player \emph{B} (blue) does not, while they share some common notes (gray).

\subsection{Improvisation Similarity}

Having many such instrument heatmaps makes comparison difficult, as there are only a few that can be shown at once without too much mental load.
We therefore explored a similarity-based layout, where more similar heatmaps are placed closer to each other using multidimensional scaling (MDS)~\cite{kruskal1964multidimensional, cutura2020druid}. 
The result is presented as a scatterplot, where each ``point'' represents a recording shown as fretboard heatmap (see above). 
Two recordings being close means that they used similar notes in the improvisation (without going into detail here on how we computer the distance).
As \autoref{fig:teaser_heatmap_dr} shows, we can use such a visualization to detect outliers, which might be interesting as they represent very different playstyles that could serve as inspiration.

The fretboard heatmap representations in this example serve as multi-dimensional glyphs~\cite{fuchs2015glyph}. 
As such, we decided to colorize the notes of each fret differently to allow for pre-attentive pattern detection. 
Saturation encodes how often a note was played (normalized over all notes).
In our example, for instance, most improvisations used the A minor pentatonic scale (a set of notes that fits the backing track well) on the fifth fret, which shows up as red-to-orange cells in the left half of \autoref{fig:teaser_heatmap_dr}.

\subsection{Music Theory for Improvisation}

In music theory, scales such as the A minor pentatonic serve as guidelines for which notes to pick and how often or long to play them.
For our blues improvisation example, there is a so-called \emph{blue note} that can be added sparingly to sound more interesting.

Based on the theory input from the guitar teacher, we thus designed an ordinal/categorical color scheme for the different roles of notes.
Our visualization then represents each played note by a rounded rectangle with the corresponding color; each line represents a different recording (\autoref{fig:teaser_scales}).
This encoding allows seeing, for instance, that the bottom most recording has a rather long blue note, which could be a mistake --- or inspiration --- to further investigate by listening to the audio.

\subsection{Discussion}

The four examples presented above are meant as illustration of the broader approach of using data visualization to further enrich instrument learning. 
In fact, we also created various other visualizations during our process, which we have not discussed here for space reasons. 
More importantly though, we believe the concept behind is much broader and can be used for many other settings, instruments, and learning/teaching approaches.

Here, we  focused on a online teacher-student group setup and guitar as the instrument of choice.
We collected plenty of feedback from our participants throughout the process.
Reporting these results in depth is beyond the scope of this paper. 
However, two of the main benefits of our approach that participants frequently mentioned were:
(1) The teacher can assess all students together, whereby subgroups of students can be defined based on skill or taste, for example, to let experienced students help less experienced ones that like to learn a common piece.
(2) Students can compare their playing and progress to others' for competitive motivation and to ask others about common issues.

In terms of visualization, we note that there are many abstractly similar types of visualizations that can be taken as inspiration. 
In our case, we use MIDI, which essentially boils down to the abstract group of temporal sequence data visualizations.
There exists a range of work on visual analysis of such temporal data~\cite{aigner2011visualization}. 
Examples include \emph{PlanningLines}~\cite{aigner2005planninglines}, \emph{LifeFlow}~\cite{Wongsuphasawat2011lifeflow}, and \emph{EventRiver}~\cite{luo2012EventRiver}. 
We used these approaches as inspiration in our design process and extended them to the domain-specific requirements of music recordings. 
Generally, we echo the value of problem and data abstraction in such design processes~\cite{munzner2009nested,meyer2015nested,sedlmair2012design}.

\section{Conclusion}

We explored various ways to leverage visualizations of recorded practice data for music education.
Instead of automatic assessment, we recommend to visually support the analysis and comparison by humans. 
A teacher can, for instance, focus time on reasoning and explanation instead of asking students how much they practiced last week.
Through a participatory design process, we designed and evaluated several means for visual comparison to others.
Feedback we gathered from our participants demonstrates that our approach helps confirm assumptions and reveal new insights.
We also found challenging design considerations, including trade-offs between user freedom and data comparability, too detailed and too abstract information, and focus on abstraction versus visual encodings.

In the future, we plan to extend our approach to other instruments, genres, and playstyles, by adapting and extending our current designs.
Furthermore, we will investigate aggregation techniques for larger datasets.
For evaluation, we plan to conduct a longitudinal field study in cooperation with a music school.
We believe that data-driven music education has potential to support teachers and students and provide new ways to practice and analyze music exercises in personal and remote learning scenarios.

\begin{acks}
Funded by Deutsche Forschungsgemeinschaft (DFG, German Research Foundation) under Germany's Excellence Strategy - EXC 2075 - 390740016, and by Cyber Valley (InstruData project).
\end{acks}

\bibliographystyle{ACM-Reference-Format}
\bibliography{sample-base}

\end{document}